\documentclass[conference]{IEEEtran}
\usepackage{cite}
\usepackage{amsmath,amssymb,amsfonts}
\usepackage{algorithmic}
\usepackage{graphicx}
\usepackage{textcomp}
\usepackage{xcolor}
\usepackage{capt-of}
\usepackage[switch]{lineno} 
\usepackage{array}
\newcommand{\PreserveBackslash}[1]{\let\temp=\\#1\let\\=\temp}
\newcolumntype{C}[1]{>{\PreserveBackslash\centering}p{#1}}
\newcolumntype{R}[1]{>{\PreserveBackslash\raggedleft}p{#1}}
\newcolumntype{L}[1]{>{\PreserveBackslash\raggedright}p{#1}}

\def\BibTeX{{\rm B\kern-.05em{\sc i\kern-.025em b}\kern-.08em
    T\kern-.1667em\lower.7ex\hbox{E}\kern-.125emX}}
\begin{document}

\title{An AI Architecture with the Capability to Classify and Explain Hardware Trojans}

\author{\IEEEauthorblockN{Paul Whitten, Francis Wolff, Chris Papachristou}
\IEEEauthorblockA{\textit{Electrical, Computer, and Systems Engineering} \\
\textit{Case School of Engineering}\\
\textit{Case Western Reserve University}\\
Cleveland, OH, USA\\
pcw@case.edu, fxw12@case.edu, cap2@case.edu}

}

\maketitle

\begin{abstract}

Hardware trojan detection methods, based on machine learning (ML) techniques,
mainly identify suspected circuits but lack the ability to explain how the
decision was arrived at. An explainable methodology and architecture is
introduced based on the existing hardware trojan detection features. Results are
provided for explaining digital hardware trojans within a netlist using
trust-hub trojan benchmarks.


\end{abstract}


\section{Introduction}

\begin{figure}
    \includegraphics[width=9.0cm]{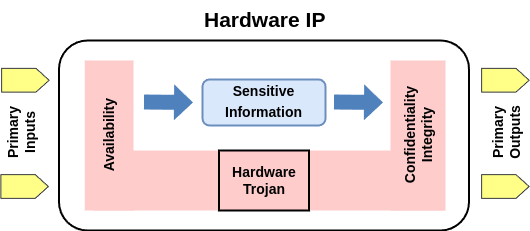}
    \caption{Hardware trojan CIA impact model.}
    \label{fig:cia_impact_model}
\end{figure}

Hardware trojans are malware circuits that are injected within an integrated
circuit (IC) during design stages, before the IC is manufactured. Once
manufactured, the trojan cannot be removed nor can it be easily bypassed by
software patches because it is baked into the IC chip. The trojan has become
part of the DNA of the IC chip and unknowingly part of the intellectual property
of the hardware design. Hence, attackers use hardware trojans to weaken the
information security of the IC chip whose properties can be summarized as
confidentiality, integrity and availability (CIA), as shown in
Fig.~\ref{fig:cia_impact_model}. For example, an availability hardware trojan
will change the privileged mode of a processor from user to supervisor mode
which allows for root access within the software operating system. An integrity
hardware trojan can periodically leak or corrupt the sensitive information, such
as encryption/decryption keys to the primary outputs (PO).

\begin{figure}
    \includegraphics[width=9.0cm]{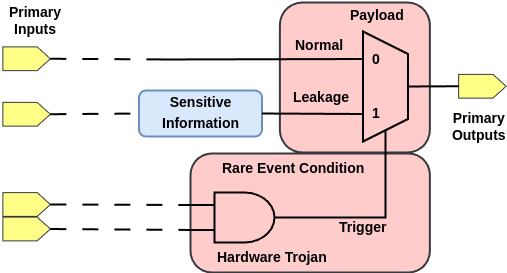}
    \caption{Confidentiality rare event hardware trojan model.}
    \label{fig:cre_impact_model}
\end{figure}

Fig.~\ref{fig:cre_impact_model} shows a specific implementation of a
confidentially hardware trojan using logic gates within an IC design file
netlist. The dashed lines represent additional levels of logic gates. The
trigger is conditional on input patterns within the IC which eventually
originate from the primary inputs. The logical-and gate requires that all the
inputs are logically true before triggering the payload. In order to be
difficult to detect or accidentally trigger during IC test mode, the hardware
trojan trigger should be a rare event condition. Given $n$ inputs to a
combinational IP circuit, the rarest condition would be one of $2^n$ inputs.
Once the hardware trojan is triggered, the multiplexor circuit within the
payload switches from normal operation to leaking sensitive information on the
primary output pin of the circuit where it can be observed by the attacker.

This rare event hardware trojan model allows for detection through static
circuit analysis by examining each net or group of logic gates within an IP
circuit primary for input or fanin complexity. The greater the fanin then the
higher the probability that a subcircuit is a potential suspect for a hardware
trojan. Furthermore, payloads need to propagate their sensitive information to
the primary output. The rare event fanin and primary output payload concepts
forms the basis of trojan analysis of several methodologies \cite{4484928,
7604700, hasegawa2020hardware, px6s-sm21-22}.

Fanin and output payload are features of a net or set of logic gates. Features
are used widely in ML classification and pattern recognition. Individual
features may contribute little to identification. The combination of features,
however, is useful for effective classification.  Combining features can also be
useful for explaining the results of a system.

The ability to effectively explain decisions is necessary to establish trust in
an automated system.  Without an explanation, a user has little understanding
and faith in a decision.  This work approaches hardware trojan identification,
in previous work, and attempts to compare two techniques of explaining
decisions.

The first method leverages combinations of features, identified as properties,
to explain decisions of a system in recognizing hardware trojans.  An
architecture is introduced that combines features, building an ML architecture
that operates on feature combinations to identify and explain decisions in
identifying hardware trojans.

The second method takes a case-based approach for explaining decisions. New
samples are related to cases the system was trained on.  A second  explainable
architecture is presented that utilizes a training index (TI) to rapidly provide
relevant examples and references to cases in netlists for context.

\section{Related Work} \label{related_work}

\begin{figure}
    \centering
    \includegraphics[width=2.5cm]{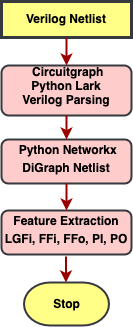}
    \caption{Data preparation flow.}
    \label{fig:data_prep}
\end{figure}

Haswega et al. introduced a method of detecting hardware trojans from a
gate-level netlist \cite{7604700, hasegawa2020hardware}. Five features from each
net are used to classify trojans.  (a) Logic Gate Fanins (LGFi) which represents
the number of inputs to the logic gates two levels upstream from a net. (b)
Flip-Flop Input (FFi) which represents the number of upstream logic levels to a
flip-flop. (c) Flip-Flop Output (FFo) which represents the number of logic
levels downstream to a flip-flop. (d) Primary Input (PI) which represents the
number of logic levels upstream to the closest primary input. (e) Primary Output
(PO) which represents the number of logic levels downstream to the closest
primary output.  In the work, the authors also used three training strategies, due
to the imbalance of the trojan to non-trojan training data ratio: no weighting,
static weighting, and dynamic weighting.  Apparently, dynamic weighting
performed best.  Other works address the imbalance in other means such as
synthetic method over-sampling techniques \cite{10444240}.  These optimized
methods still do not lead to a better understanding of the ML results.

Several works discuss the combination of results of multiple trained neural
networks (NN). Jacobs, et al. identified local experts of trained networks
\cite{6797059}. Other works treated multiple trained networks as committees and
combined NN to form a collective decision
\cite{perrone1993putting, sharkey1996combining}.

An explainable additive NN model is posed by Vaughan et al.
where a system is composed layering distinct NNs that are trained on transforms
of the inputs.  A layer then combines the outputs of the distinct NNs to perform
a prediction. Explainability comes from each distinct NN modeling features of
the input which lends to interpretability of the architecture
\cite{vaughan2018explainable}.

An explainable Artificial Intelligence architecture (XAI) using properties,
transforms of input related to the properties, Inference Engines (IE) for each
property, probabilistic decision-making, and attributing decisions to relevant
explainable properties was posed\cite{whitten21, whitten23, whitten24}.  Like
combined NN and additive NN systems, the explainable architecture examined
decisions of multiple NNs.

Case-based explanations for medical models, introduced by Caruana et al.,
suggested using a method based on k-nearest neighbor (KNN) distance in
multidimensional feature space as effective in identifying like cases from
training as explanations for new samples \cite{Caruana1999CasebasedEO}. Like
cases from training should produce similar results to new samples.  The
case-based method further suggested that leveraging training data is more
difficult for more complex models such as NN as the training set is discarded.
In the case of NNs, the activation of $n$ hidden neurons are translated into
points in an $n$-dimensional space and a KNN algorithm can be applied to find
similar activation patterns. While this may suggest similarity to the NN's
activation and behavior between like cases, the method does little to explain
what is going on in the NN.


\section{Method} \label{method}

\subsection{Data Preparation}

This work utilized trust-hub.org trojan benchmark data \cite{6657085,
shakya2017benchmarking, px6s-sm21-22, slayback2015computer}.  The particular
fifteen netlists used from trust-hub were: RS232-T1000, RS232-T1100,
RS232-T1200, RS232-T1300, RS232-T1400, RS232-T1500, RS232-T1600, s15850-T100,
s35932-T100, s35932-T200, s35932-T300, s38417-T100, s38417-T200, s38584-T100,
and s38584-T300.

Fig.~\ref{fig:data_prep} depicts the flow of data processing and preparation.
The verilog gate-level netlists containing trojans were processed using
circuitgraph \cite{circuitgraph} which uses Lark, a Python parsing toolkit
\cite{lark}.  Directed graph representations of the netlists were built in
NetworkX \cite{networkx} and then queried to obtain the five features and class
for each gate. Class can be defined as trojan (1) or non-trojan (0).  Herein,
when referencing class, trojan is often abbreviated as $t$ and non trojan as $n$.
Submodules in trust-hub netlists were not processed. Omitting netlist submodules
reduced the number of trojans in some netlists compared to other works.

The focus of this work is explainability. Rather than consider each unique
netlist as a test set and the remaining netlists as a training set, all netlists
were combined.  20\% of the combined 52,737 entries from the fifteen
netlists were sampled at random and saved as a test set. The remaining (80\%)
samples were used as a training set.

\subsection{Property-Based Explainable Architecture} \label{prop_explainable_arch}



\begin{figure}
    \includegraphics[width=9.0cm]{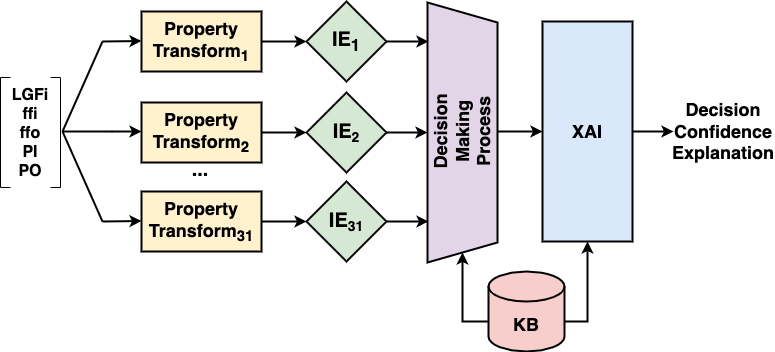}
    \caption{A property-based explainable architecture for trojan detection.}
    \label{fig:trojan_arch}
\end{figure}



Fig.~\ref{fig:trojan_arch} depicts an explainable architecture for
classification of hardware trojans. The five features (LGFi, FFi, FFo, PI, PO)
are shown as input to the architecture. The five features are transformed into
all possible 31 combinations of features known as properties depicted in Table
\ref{tab_prop_id}. The ID columns are the identifiers, $j$, for each property.
The features columns list the feature combination for that property.  The
$X_j$ columns denote the property's explainability.

Each of the properties has an Inference Engine (IE) shown in
Fig.~\ref{fig:trojan_arch} that is used to cast votes based on the property. IEs
were implemented as Support Vector Machines (SVM) from Scikit-learn
\cite{scikit-learn}. The SVMs used a radial basis function kernel.
SVM $C$ and $\gamma$ parameters were optimally set and then SVMs were trained
based on training data.

\begin{table*}
    \renewcommand{\arraystretch}{1.3}
    \centering
    \caption{Properties, Features, and Explainability}
    \resizebox{1.05\columnwidth}{!}{%
    \begin{tabular}{|c|c|c||c|c|c||c|c|c|}
        \hline
        \bfseries ID & \bfseries Features & \bfseries $X_j$ & \bfseries ID & \bfseries Features & \bfseries $X_j$ & \bfseries ID & \bfseries Features & \bfseries $X_j$ \\
        \hline
        \hline
        1 & LGFI & 1.00 & 12 & ffi, PO & 0.75 & 23 & ffi, ffo, PO & 0.50 \\
        \hline
        2 & ffi & 1.00 & 13 & ffo, PI  & 0.75 & 24 & ffi, PI, PO & 0.50 \\
        \hline
        3 & ffo & 1.00 & 14 & ffo, PO & 0.75 & 25 & ffo, PI, PO & 0.50 \\
        \hline
        4 & PI & 1.00 & 15 & PI, PO & 0.75 & 26 & LGFi, ffi, ffo, PI & 0.25 \\
        \hline
        5 & PO & 1.00 & 16 & LGFi, ffi, ffo & 0.50 & 27 & LGFi, ffi, ffo, PO & 0.25 \\
        \hline
        6 & LGFi, ffi & 0.75 & 17 &  LGFi, ffi, PI & 0.50 & 28 & LGFi, ffi, PI, PO & 0.25 \\
        \hline
        7 & LGFi, ffo & 0.75 & 18 & LGFi, ffi, PO & 0.50 & 29 & LGFi, ffo, PI, PO & 0.25 \\
        \hline
        8 & LGFi, PI & 0.75 & 19 & LGFi, ffo, PI & 0.50 & 30 & ffi, ffo, PI, PO & 0.25 \\
        \hline
        9 & LGFi, PO & 0.75 & 20 & LGFi, ffo, PO & 0.50 & 31 & LGFi, ffi, ffo, PI, PO & 0.0\\
        \hline
        10 & ffi, ffo & 0.75 & 21 & LGFi, PI, PO & 0.50 \\
        \cline{0-5}
        11 & ffi, PI & 0.75 & 22 & ffi, ffo, PI & 0.50 \\
        \cline{0-5}
    \end{tabular}
    }
    \label{tab_prop_id}
\end{table*}

During implementation of the IEs, no weighting, static, and dynamic weighting
methods were compared.  Static and dynamic weighting far outperformed no
weighting. Both static weighting and dynamic weighting help overcome the highly
imbalanced dataset.  When the static method utilizes a balance factor, $b$,
given in \ref{eq:balance}, rather then an arbitrary number, it performs
comparably to dynamic weighting.  The balance factor, $b$, is given by the
number of non-trojan nets over the number of trojan nets.  In the 80\% training
set, $b(trojan)=\frac{42190}{160}=263.7$.  When training the architecture, the
static method, using balance, was employed due its performance and simplicity.


\begin{equation}\label{eq:balance}
    b(class) =
    \begin{cases}
        \frac{|non-trojan|}{|trojan|},& class = \text{ trojan ($t$)} \\
        1.0,& class = \text{non-trojan ($n$)}
    \end{cases}
\end{equation}

The Decision Making Process (DMP) is responsible for considering the votes from
the IEs and making a system-wide decision. As the DMP receives votes from the
IEs, it obtains property effectiveness weights from the KB. Effectiveness of the
properties was stored in the KB by reprocessing the training data and analyzing
results to obtain performance metrics for each IE. Effectiveness as weights are
used in tallying the votes to produce decisions, suspected trojan ($t$) or
non-trojan ($n$), and confidence of the decisions.  As was experienced with
other highly imbalanced datasets, measuring effectiveness is a challenge.
Effectiveness metrics were tried and $E_{PARS}$ performed best \cite{whitten24}.

The XAI function takes the decisions, votes, and confidence producing an
explanation to justify the decisions of the system.  The explanations relate to
properties, hinting at relationships between features.

In building the rationale in XAI, a threshold of 0.05 was used for registering a
vote worthy of mention.  Without this threshold, each of the 31 properties were
listed in the rationale, many with very little weight.  The rationale composed
by XAI is also organized with weights of properties sorted in a descending
fashion to mention the highest contributing properties first.


Each property $P_j$ has an explainability metric, $X_j$, signifying how explainable
that property is.  Explainability, $X_j$, is given in \eqref{eq:explainability}.
The Explainability metric is based upon the number of features or cardinality of
the $j^{th}$ property, $P$.  Again, $n$, is the number of features in the input
vector to the architecture.

A property with one feature would have high explainability, $X_j=1.0$.
Properties with few features are more explainable, while a property with more
features is difficult to explain.  Intuitively, this corresponds with reasoning
that a ML model with the five features used by Haswega acted as an opaque box.
If a single feature could easily indicate a trojan, that is likely explainable
to a user.  Explainability for property combinations in this work for the
properties is indicated in the $X_j$ column of Table \ref{tab_prop_id}.

\begin{equation}\label{eq:explainability}
    X_j = 1.0 - \frac{|P_j| - 1}{n - 1}
\end{equation}


\subsection{Case-Based Explainable Architecture} \label{case_explainable_arch}

\begin{figure}
    \centering
    \includegraphics[width=8.0cm]{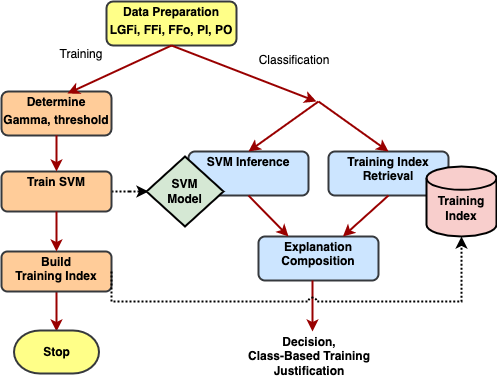}
    \caption{Flow of the case-based explainable method.}
    \label{fig:case_xai_flow}
\end{figure}


\begin{table}
    \renewcommand{\arraystretch}{1.3}
    \centering
    \caption{Examples from the test set}
    \begin{tabular}{|c|c|c|c|c|c|c|}
        \hline
        \bfseries Example ID & \bfseries LGFi & \bfseries FFi & \bfseries FFo & \bfseries PI & \bfseries PO & \bfseries Trojan \\
        \hline
        \hline
        1 & 8 & 1 & 3 & 2 & 3 & 1 \\
        \hline
        2 & 3 & 3 & 5 & 3 & 5 & 1 \\
        \hline
        3 & 5 & 2 & 14 & 2 & 14 & 0 \\
        \hline
    \end{tabular}
    \label{tab_ex}
\end{table}

\begin{figure}
    \centering
    \includegraphics[width=7.0cm]{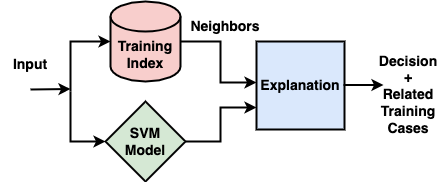}
    \caption{A case-based explainable architecture using a training index.}
    \label{fig:case_xai_arch}
\end{figure}

SVMs are based on fitting hyperplanes to bound, in multidimensional space,
classes that are learned by training data.  KNN and the concept of distance
translate well to the multidimensional space of features.  Despite the notion
suggested by Caruana et al., that KNN between a new sample and training data
only explains the training data, this work suggests that using a KNN approach to
training data is effective for providing an explanation, backed by training
cases.

Fig.~\ref{fig:case_xai_flow} depicts the flow of the case-based explainable
method. It starts from the features from data preparation.  The initial training
phase, shown on the left vertical flow, includes determining $\gamma$ and $C$
for the SVM.  Training the SVM using the training data is completed next to
produce an SVM model. Lastly, a training index is built.  The right portion of
the flow in Fig.~\ref{fig:case_xai_flow} illustrates classification and
explanation of new samples using the SVM Model and TI created in training.  The
decision from the SVM model and case-based justification from TI serve as input
to compose results provided to the user.

A case-based explainable architecture is presented in
Fig.~\ref{fig:case_xai_arch} using the trained SVM model, a TI, and an
explanation component from the flow.  Input samples are fed in at the left as a
feature vector.  The trained SVM model contributes a decision to the system. The
TI provides efficient retrieval of cases, supporting data used to justify the
decision.  The explanation component organizes the decision and KNN supporting
data to provide justification for the decision, in the form of neighbor
information from the training set and references to the training samples in
context.

KNN information presented to the user includes distance from the sample being
considered, the classes (trojan and non-trojan) of training samples, and
references to the samples in the context of the netlists they were extracted
from.  In addition to this information, the weight of neighboring trojans,
$w(t)$, and non-trojans, $w(n)$, are calculated relative to the inverse square
of distance, $d$, as shown in \ref{eq:weight}.  Balance, from \ref{eq:balance},
is summed for each set of training samples at the distance to the sample
considered.  The weights are then summed to provide a correspondence metric,
$C$, for KNN as shown in \ref{eq:correspondence}.

\begin{equation}\label{eq:weight}
    w(class) = \sum_{i=1}^{|class|} \frac{b(class)}{(d_i+1.0)^2} 
\end{equation}

\begin{equation}\label{eq:correspondence}
    C(class) = \frac{\sum_{i=1}^{k}{w(class)}}{\sum_{j=1}^{classes}{w(j)}} 
\end{equation}

An important property of data preparation for explainability and trust in
case-based explanations is preserving the origin of each sample for storage in
the TI as the samples are combined for training.  The origin in the context of
training samples is comprised of the part number, version (indicating which of
the 15 circuits the sample came from), line, name, and net.  While originating
fields are not useful to decision-making and classification, they provide
crucial context for the explanation.

The TI was constructed of in-memory hashmaps to quickly retrieve the supporting
data based on features. Hashmaps worked well for this application and relatively
small dataset.  In larger datasets, other key-value stores could be employed
with comparable results.  Because of the small size of the dataset and current
compute, a simple linear search was sufficient for identifying nearest neighbors
from training. As the dataset scales, a variety of more efficient algorithms
exist for searching for nearest neighbors \cite{615448, abbasifard2014survey}.

\section{Results} \label{results}

SVMs were trained using the three methods: no weighting, static weighting, and
dynamic weighting.  When static weighting used a balance $b$ corresponding to
the ratio of non-trojans to trojans, static weighting appeared to perform
equivalent to dynamic weighting.




Table~\ref{tab_ex} depicts the three examples from the test set used to
demonstrate explainability. The first column in the table represents the example
identifier (ID). The next five columns are the features.  The last column
indicates if the sample is a trojan (1) or non-trojan (0).

\subsection{Property Based Explainable Results} \label{prop_exp_results}



Table~\ref{tab_ex} displays examples from the test set (20\%) used to exhibit
explainability. This section will present the property-based architecture
results from examples using the static weighting technique.

\begin{table}
    \renewcommand{\arraystretch}{1.3}
    \centering
    \caption{Example One Property-Based Results}
    \begin{tabular}{|c|c|p{0.32\columnwidth}|c|}
        \hline
        \bfseries Prediction & \bfseries Confidence & \bfseries Properties (Table I) & \bfseries Explainability \\
        \hline
        \hline
        1 & 99.6\% & 31, 28, 26, 29, 27, 30, 21, 18, 16, 19, 24, 17, 22, 25, 15 & 41.1\% \\
        \hline
        0 & 0.4\% & No opinion & 0\% \\
        \hline
    \end{tabular}
    \label{tab_prop_exp_ex_1}
\end{table}

Example one, as shown in Table~\ref{tab_prop_exp_ex_1}, was appropriately
predicted as a trojan (1) with 99.6\% confidence.  Fifteen properties, listed by
ID from Table~\ref{tab_prop_id} in order of effectiveness, contributed
sufficiently weighted votes to be registered (above the 5\% threshold).  Despite
the high confidence, explainability was weighted at 41.1\%, because fourteen of
the properties above the threshold had three or more features, so they had
limited explainability ($X_j\leq0.5$ in Table \ref{tab_prop_id}). The alternate
decision to identify example one as a trojan had confidence $<$ 1\% with none of
the remaining properties above the threshold.


\begin{table}
    \renewcommand{\arraystretch}{1.3}
    \centering
    \caption{Example Two Property-Based Results}
    \begin{tabular}{|c|c|p{0.32\columnwidth}|c|}
        \hline
        \bfseries Prediction & \bfseries Confidence & \bfseries Properties (Table I) & \bfseries Explainability \\
        \hline
        \hline
        1 & 25.7\% & 30, 24, 22, 25, 15 & 50.0\% \\
        \hline
        0 & 74.3\% & 31, 28, 26, 29, 27, 21, 18, 16, 19, 17 & 35.0\% \\
        \hline
    \end{tabular}
    \label{tab_prop_exp_ex_2}
\end{table}

The property-based architecture improperly identified example two, suggesting
the winning decision with medium, 74.3\%, confidence was non-trojan.
Table~\ref{tab_prop_exp_ex_2} illustrates the properties, confidence, and
explainability for example two.  Ten properties voted for a non-trojan, while
five suggested a trojan.  Again, the explainability was low at only 35.0\% for
the winning decision due to the relatively large number of features in voting
properties.



\begin{table}
    \renewcommand{\arraystretch}{1.3}
    \centering
    \caption{Example Three Property-Based Results}
    \begin{tabular}{|c|c|p{0.32\columnwidth}|c|}
        \hline
        \bfseries Prediction & \bfseries Confidence & \bfseries Properties (Table I) & \bfseries Explainability \\
        \hline
        \hline
        1 & 58.4\% & 28, 26, 21, 18, 16, 20, 24, 22, 15 & 47.2\% \\
        \hline
        0 & 41.6\% & 31, 29, 27, 30, 17, 24 & 29.1\% \\
        \hline
    \end{tabular}
    \label{tab_prop_exp_ex_3}
\end{table}

Example three is labeled as non-trojan in the test set.  The architecture
improperly identified the sample as a trojan with medium, 58.4\%, confidence. In
Table~\ref{tab_prop_exp_ex_3} nine properties suggested the sample as a trojan while
six suggested a normal net.  Explainability was among the highest at 47.2\%.



\subsection{Case-Based Explainable Results} \label{case_exp_results}

The case based explainable architecture performed well, providing KNN for all
of the test samples.  Weights of KNN corresponded to the decision of the
architecture in 97.4\% of the cases.


\begin{table}
    \renewcommand{\arraystretch}{1.3}
    \centering
    \caption{Example One Case-Based Results}
    \resizebox{\columnwidth}{!}{%
    \begin{tabular}{|c|c|c|r|r|}
        \hline
        \bfseries Distance & \bfseries Feature Values & \boldmath$t:n$ & \boldmath$w(t)$ & \boldmath$w(n)$ \\
        \hline
        \hline
        0.00 & \textlangle 8, 1, 3, 2, 3\textrangle & 11:0 & 2893.0 & 0.00 \\
        \hline
        1.00 & \textlangle 8, 2, 3, 2, 3\textrangle & 0:34 & 0.00 & 8.50 \\
        \hline
        1.00 & \textlangle 8, 1, 3, 1, 3\textrangle & 1:204 & 65.75 & 51.00 \\
        \hline
        1.41 & \textlangle 8, 1, 4, 2, 4\textrangle & 3:0 & 135.37 & 0.00 \\
        \hline
        \cline{2-5}
        \multicolumn{1}{c|}{} & \multicolumn{1}{l|}{\textbf{Sum}} & $15+238=253$ & 3094.12 & 59.50 \\
        \cline{2-5}
        \multicolumn{2}{c|}{} & \multicolumn{1}{l|}{\textbf{Correspondence}} & 98.1\% & 1.9\% \\
        \cline{3-5}
    \end{tabular}
    }
    \label{tab_index_exp_ex_1}
\end{table}


The SVN in the case-based architecture for example one voted for a trojan.  The
KNN cases provided by the architecture are shown in
Table~\ref{tab_index_exp_ex_1} where the distance column indicates the euclidean
distance to the sample.  The feature values indicate the feature vector of the
training sample(s).  The $t:n$ column indicates the ratio of trojan ($t$) to
non-trojan ($n$) nets for the samples in the training set. The $w(class)$
columns indicate the weight for each class, trojan ($t$) and non-trojan ($n$)
from \ref{eq:weight}.

The case-based architecture further provides a quantitative measure for example
one indicating the decision corresponds with neighbor weighting by suggesting a
correspondence of 98.1\% for the prediction of trojan based on neighbors from
training data.

References to the neighboring cases for example one are helpful in identifying
those cases in context.  Neighbors at \textlangle 8, 1, 3, 2, 3\textrangle
include all eleven matches e.g., part number RS232, version T1500, line 40, name
NAND4X1, and net U294.QN as well as part RS232, version T1000, line 35, name
NAND4X1, and net U299.QN. The output of the explainability results produces all
references to KNN examples from the case-based training set.


\begin{table}
    \renewcommand{\arraystretch}{1.3}
    \centering
    \caption{Example Two Case-Based Results}
    \begin{tabular}{|c|c|c|r|r|}
        \hline
        \bfseries Distance & \bfseries Feature Values & \boldmath$t:n$ & \boldmath$w(t)$ & \boldmath$w(n)$ \\
        \hline
        \hline
        0.00 & \textlangle 3, 3, 5, 3, 5\textrangle & 0:57 & 0.00 & 57.00 \\
        \hline
        1.00 & \textlangle 4, 3, 5, 3, 5\textrangle & 1:24 & 65.75 & 6.00 \\
        \hline
        1.00 & \textlangle 2, 3, 5, 3, 5\textrangle & 0:104 & 0.00 & 26.00 \\
        \hline
        2.00 & \textlangle 5, 3, 5, 3, 5\textrangle & 0:11 & 0.00 & 1.22 \\
        \hline
        \cline{2-5}
        \multicolumn{1}{c|}{} & \multicolumn{1}{l|}{\textbf{Sum}} & $1+196=197$ & 65.75 & 90.22 \\
        \cline{2-5}
        \multicolumn{2}{c|}{} & \multicolumn{1}{l|}{\textbf{Correspondence}} & 42.2\% & 57.8\% \\
        \cline{3-5}
    \end{tabular}
    \label{tab_index_exp_ex_2}
\end{table}

Example two was incorrectly identified as non-trojan by the case-based
architecture. The KNN cases are depicted in Table~\ref{tab_index_exp_ex_2}. The
architecture assigned a correspondence of 57.8\% for the prediction of no trojan
based on neighbors from training data. This was due to the single neighboring
trojan's weight at distance one being overcome by other neighboring 196 samples.
References to all 197 samples from training were provided for context.


\begin{table}
    \renewcommand{\arraystretch}{1.3}
    \centering
    \caption{Example Three Case-Based Results}
    \begin{tabular}{|c|c|c|r|r|}
        \hline
        \bfseries Distance & \bfseries Feature Values & \boldmath$t:n$ & \boldmath$w(t)$ & \boldmath$w(n)$ \\
        \hline
        \hline
        0.00 & \textlangle 5, 2, 14, 2, 14\textrangle & 0:1 & 0.00 & 1.00 \\
        \hline
        1.41 & \textlangle 5, 2, 13, 2, 13\textrangle & 0:2 & 0.00 & 0.34 \\
        \hline
        1.73 & \textlangle 4, 2, 13, 2, 13\textrangle & 7:0 & 246.65 & 0.00 \\
        \hline
        2.45 & \textlangle 3, 2, 13, 2, 13\textrangle & 3:0 & 66.31 & 0.0 \\
        \hline
        \cline{2-5}
        \multicolumn{1}{c|}{} & \multicolumn{1}{l|}{\textbf{Sum}} & $10+3=13$ & 312.96 & 1.34 \\
        \cline{2-5}
        \multicolumn{2}{c|}{} & \multicolumn{1}{l|}{\textbf{Correspondence}} & 99.6\% & 0.4\% \\
        \cline{3-5}
    \end{tabular}
    \label{tab_index_exp_ex_3}
\end{table}

Example three, shown in Table~\ref{tab_index_exp_ex_3}, was correctly identified
as not a trojan by the SVM.  The KNN gave a rare conflicting 99.6\%
correspondence metric for a trojan contradicting the prediction of the SVM.  The
two closest samples from training were all non-trojan with three samples.  The
three samples were overcome by the ten samples that were the next closest due to
the balance.




\section{Conclusion}

The property-based explainable architecture produced rationale for decisions
that had only marginal utility in relating the decisions to combinations of
features of the input data.  Explainability metrics for the property-based
architecture were fairly low for the properties due to a generally high number
of features in the properties above the threshold.  The low explainability
indicates that the rationale for the property based decisions was not
compelling. As previously observed, with the property-based architecture,
explainability was at the cost of accuracy compared to other leading
unexplainable methods.


Explanations and justification for the case-based explainable architecture
outperformed the property based architecture on a subjective basis.  The
case-based architecture also benefitted from the slightly more accurate but
unexplainable SVM. The cases from training were relevant and references to
originating netlists of training samples added credibility and trust to the
system.  Quantitatively, the case-based architecture provided high
correspondence, 97.4\%, in weighting neighboring training cases to the decisions
of the SVM.

Attempting to apply the property-based explainable architecture to a dataset
with a low dimensional space (five features) did not work as well as a
case-based architecture in explaining results in the case of hardware trojan
detection.  The case-based architecture clearly outperformed the property-based
architecture in providing explanations and establishing trust in this
application.



\bibliographystyle{IEEEtran}
\begin{flushleft}
\bibliography{references}
\end{flushleft}

\end{document}